# Electronic states and magnetic structure at the $Co_3O_4$ (110) surface: a first principles study

*Jia Chen, Annabella Selloni*

Department of Chemistry, Princeton University, Princeton, New Jersey 08544, USA

ABSTRACT

Tricobalt tetraoxide ($Co_3O_4$) is an important catalyst and $Co_3O_4$(110) is a frequently exposed surface in $Co_3O_4$ nanomaterials. We employed Density-functional theory with on-site Coulomb repulsion U term to study the atomic structures, energetics, magnetic and electronic properties of the two possible terminations, A and B, of this surface. These calculations predict A as the stable termination in a wide range of oxygen chemical potentials, consistent with recent experimental observations. The $Co^{3+}$ ions do not have a magnetic moment in the bulk, but become magnetic at the surface, which leads to surface magnetic orderings different from the one in the bulk. Surface electronic states are present in the lower half of the bulk band gap and cause partial metallization of both surface terminations. These states are responsible for the charge compensation mechanism stabilizing both polar terminations. The computed critical thickness for polarity compensation is 4 layers.

# I. INTRODUCTION

The spinel cobalt oxide $Co_3O_4$ is a magnetic semiconductor and widely used catalyst for a variety of reactions.[1,2] Recently, this material has attracted further interest as a promising catalyst for energy and environment-related applications such as low-temperature CO oxidation,[3] water splitting,[4] and the oxygen reduction reaction.[5] Surfaces have a key role in these applications, and a detailed understanding of the physical and chemical properties of $Co_3O_4$ surfaces is important for the design of $Co_3O_4$-based functional materials with improved performance. Experimental atomic-scale investigations of $Co_3O_4$ surfaces are relatively scarce however. To help obtaining a better fundamental understanding of the surface properties of $Co_3O_4$, in this work we present a first principles Density Functional Theory (DFT) study of the $Co_3O_4$(110) surface, which is the predominant one on $Co_3O_4$ nano-rods,[3] and is believed to be mainly responsible for the oxidation reactivity[6] of this material.

$Co_3O_4$ crystallizes in the cubic normal spinel structure with magnetic $Co^{2+}$ ions in tetrahedral sites and non-magnetic $Co^{3+}$ ions in octahedral sites. The (110) surface is a Type III polar surface according to Tasker's criterion.[7] It has two different terminations, usually denoted as the A and B terminations (see Fig. 1): the (110)-A termination exposes both $Co^{2+}$ and $Co^{3+}$ ions, whereas the (110)-B termination has only $Co^{3+}$ ions. As $Co_3O_4$ is basically ionic,[8] the unit cell of the (110)-A termination − exposing two $Co^{2+}$, two $Co^{3+}$, and four $O^{2-}$ ions − has formal charge +2, whereas the same unit cell on the (110)-B termination exposes two $Co^{3+}$ and four $O^{2-}$ ions, and therefore has formal charge -2. Thus a (110) slab can be viewed as a stack of charged layers as sketched in Fig. 2. While in principle such a system has a polarization which increases linearly with slab thickness and eventually diverges, in reality polarity compensation mechanisms exist which prevent the "polar catastrophe" and stabilize the surface.[9] (also see Fig. 2)

A number of first principles studies of $Co_3O_4$(110) have been already reported,[10-13] but some basic properties, including the polarity compensation mechanism, have not been examined in detail and/or are not yet well understood.[14,15] An objective of this work is thus to investigate how polarity is compensated on the two different surface terminations of $Co_3O_4$(110). Since experiments do not show evidence of surface reconstruction on either termination,[14,15] we will restrict to undefected and unreconstructed (110)-A and (110)-B terminations obtained by simply relaxing the bulk truncated structures, and will study the compensation mechanism by focusing on the surface electronic structure. We will also examine the surface magnetic structure as recent experiments on $Co_3O_4$ nanostructures[16-20] have revealed interesting features which cannot be fully explained simply on the basis of the magnetic properties of bulk $Co_3O_4$.

Following our recent investigation of bulk $Co_3O_4$,[8] the present study of the $Co_3O_4$(110) surface is based on DFT calculations within the generalized gradient approximation (GGA) [14,15] augmented with an on-site Coulomb repulsion U term in the 3d shell of the cobalt ions. The GGA+U approach reduces significantly the delocalization error arising from the incomplete cancellation of the Coulomb self-interaction in pure GGA calculations,[21] and gives a value of the band gap for bulk $Co_3O_4$ (1.96 eV) in reasonable agreement with experiment (~1.6 eV).[22,23] The U repulsion terms in Ref. [8] were determined from first principles using linear response.[24] The resulting values, U = 4.4 and 6.7 eV for the $Co^{2+}$ and $Co^{3+}$ ions, respectively, reflect the different oxidation states and local electronic structure of the two ions. For surfaces, however, it is difficult to pre-identify the oxidation states of the surface Co ions. Moreover, the use of multiple U values renders the calculation of surface energies and other thermodynamics quantities more involved. Therefore in this work we use a single U value for all Co ions in our models, namely U = 5.9 eV, which corresponds to the weighted average of the two computed U values for the bulk. The bulk properties computed using this U for all Co ions are very similar to those reported in Ref. [8]. For example, the band gap is 1.96eV using two U values, and 1.92eV using their weighted average.

This paper is organized as follows. After a brief description of the computational methods in Sec. II, in Sec. III we first present our results on the surface structural, magnetic, and electronic properties. Next, based on analysis of the Wannier functions, the polarity compensation and surface charge are discussed, and the critical thickness for polarity compensation is evaluated. Conclusions are given in Sec. IV.

## II. METHODS AND MODELS

Calculations were performed within the plane wave-pseudopotential scheme as implemented in the Quantum Espresso package.[25] Spin polarization was always included and exchange and correlation were described using the gradient corrected Perdew-Burke-Ernzerhof (PBE)[26] functional with on-site Coulomb repulsion U term on the Co 3d states. As mentioned in the Introduction, we used a single U value for all Co ions, namely U = 5.9 eV, which corresponds to the weighted average of the two computed U values for the bulk.[8] For comparison, pure PBE calculations have been also performed; however, unless otherwise specified, only PBE+U results are reported in the following. Ultrasoft pseudopotentials[27] were used and the valence electrons included O 2s, 2p and Co 3d, 4s states. Plane wave energy cutoffs of 35 Ry for the smooth part of the wavefunction and 350 Ry for the augmented density were found sufficient to ensure a good convergence of the computed properties.

Surfaces were modeled using a periodic slab geometry, with consecutive slabs separated by a vacuum layer 15 Å wide. We adopted the PBE+U lattice constant from our previous work which is 2% larger than the experimental one.[8] (Pure PBE calculations were performed with the corresponding optimized lattice constant.[8]) To study the properties of a single A or B termination, we considered symmetric slabs with odd number of layers, for which the total dipole moment is zero. Although non stoichiometric, these models provide useful information in the thick sample limit, when the effect of the nonstoichiometry becomes negligible.[28] We performed tests on slabs with different number of layers, from 5 up to 11 layers, and found that a well converged description could be achieved with 9-layer models. On the other hand, to achieve perfect stoichiometry, one should consider slabs with even number of layers, which expose the A and B terminations on the two different sides, and have a dipole moment perpendicular to the slab. We also performed tests to compare the results obtained with symmetric and non-symmetric slabs and found that the surface properties (e.g. the surface electronic structures of the different terminations, see Sect. 3C) obtained with 9-layer models agree well with those from symmetric slabs of 8 or 10 layers. Results reported in the following thus refer to calculations on 9-layer models, unless otherwise specified. Structural optimizations were carried out by relaxing all atomic positions until all forces were smaller than $1\times10^{-3}$ a.u.

For most calculations the rectangular surface cell depicted in Figure 1 was used, and the sampling of the surface Brillouin zone was performed using a 3×4 k-point grid. Comparisons to calculations using a 4×6 k-point grid show surface energy differences of ~ 1meV/Å$^2$. Maximally-localized Wannier functions (MLWFs)[29] were obtained using the $\Gamma$ point only on models with a surface supercell twice the size of the rectangular cell in Fig. 1. Test calculations showed that the results for the two setups were in satisfactory agreement. The MLWFs were calculated with the algorithm developed by Sharma et al.[30].

# III RESULTS AND DISCUSSION

## A. Energetics and structure

### A.1. Surface energies

Experimental studies on $Co_3O_4$(110) epitaxial films grown on $MgAl_2O_4$(110) single crystal substrates found that the surfaces of the as-grown films are relatively disordered and have an oblique low-energy electron diffraction (LEED) pattern characteristic of the (110)-B termination, whereas the annealed surfaces show a sharp rectangular LEED pattern indicating a well ordered (110)-A termination.[14,15] These findings indicate that the (110)-A termination is more stable than the B one under Ultra High Vacuum (UHV) conditions. However, the occurrence of the (110)-B termination on the as-grown films suggests the existence of kinetic limitations,[14] so that the actual exposed termination may depend on the synthetic method and the post-treatment of the samples.

In order to study the properties of a single termination, it is convenient to consider symmetric, non-stoichiometric slabs, and express their surface formation energies in terms of the chemical potentials of Co ($\mu_{Co}$) and oxygen ($\mu_O$).[31] Since $3\mu_{Co} + 4\mu_O = \mu_{Co3O4}$ under equilibrium conditions, $\mu_{Co3O4}$ being the chemical potential of bulk $Co_3O_4$, it is possible to eliminate the dependence on $\mu_{Co}$, and express the surface energy only in terms of the oxygen chemical potential $\mu_O$ or, equivalently, $\mu_O' \equiv \mu_O - \frac{1}{2} E_{tot}(O_2)$, where $E_{tot}(O_2)$ is the total energy of an $O_2$ molecule. The oxygen potential $\mu_O'$ satisfies the condition $1/4\ H_f \leq \mu_O' \leq 0$, where $H_f$ is the heat of formation of bulk $Co_3O_4$ and the lower and upper limits correspond to O-poor and O-rich conditions, respectively. Values for $H_f$ are given in Ref. [8].

The computed surface energies for slab models with 5, 7 and 9 layers in the O-rich limit ($\mu_O' = 0$) are listed in Table 1, whereas Fig. 3 shows the surface energies in the full range of $\mu_O'$ for the 9-layer slabs. For the sake of comparison with previous GGA calculations,[13] in Figure 3 results obtained at both the pure PBE and PBE+U levels are presented. We can see a significant difference between the results of the two approaches. According to the pure PBE calculations the (110)-B termination has lower surface energy except at very low $\mu_O'$, in agreement with previous published results.[13] By contrast, the PBE+U calculations predict the A-(110) termination to be more stable in a wide range of the oxygen chemical potential, consistent with the experimental results of Ref. [14]. This difference between the PBE and PBE+U results can be understood on the basis of the computed surface electronic structures, reported in Sect. 3C. Briefly, the B termination is found to have delocalized metallic surface states, for which the energy penalty from the Hubbard U term is larger, thus making the surface energy of the B termination higher. The PBE functional is known to overestimate the $O_2$ binding energy:[26] our computed value is 130 Kcal/mol, against 118 kcal/mol from experiment. This error affects the chemical potential of the oxygen rich limit as indicated in Fig. 3.

**Table 1.** Surface energies of $Co_3O_4$(110), computed at the PBE+U level and in the O-rich limit, for symmetric slabs of different thicknesses.

|  | Surface Energy (eV/Å$^2$) | |
| --- | --- | --- |
|  | A termination | B termination |
| 5-layer | 0.081 | 0.080 |
| 7-layer | 0.085 | 0.081 |
| 9-layer | 0.082 | 0.080 |

*A.2. Surface relaxation*

The A-terminated $Co_3O_4$(110) surface exposes all types of ions present in the bulk, namely $Co^{2+}$, $Co^{3+}$ and $O^{2-}$ ions. (We identify the surface ions with the oxidation state they have in the bulk, even though their actual oxidation state may be different at the surface.) The $Co^{2+}$ ($Co^{3+}$) ions are 3-fold (4-fold) coordinated and form bonds with two surface oxygen ions and one (two) oxygen(s) in the second layer; they will be denoted Co-3f (Co-4f) in the following. All surface oxygens are equivalent and 3-fold coordinated to one Co-3f and one Co-4f surface ion as well as to one 6-fold $Co^{3+}$ in the second layer (see Fig. 1). Calculated atomic relaxations on the (110)-A termination are listed in Table 2. While all surface atoms undergo an inward relaxation, this relaxation is larger for the Co than for the oxygen ions, and therefore the surface becomes slightly buckled. The reflection symmetry of the surface remains during relaxation, so that on the relaxed (110)-A surface there is one type of 3-fold and one type of 4-fold Co ion as well as one type of oxygen ion. As shown in Table 2, all surface Co-O bonds are shorter after relaxation.

**Table 2.** Atomic displacements from bulk-like positions on the relaxed (110)-A surface. Displacements along the [001], [1$\bar{1}$0] and [110] directions are denoted as (Δx, Δy, Δz). Atoms are labeled as in Fig.1.

|  | Atomic displacement (Å) | | | Bond expansion | |
|---|---|---|---|---|---|
|  | Δx | Δy | Δz | Label | Δ |
| Co3f1 | 0.17 | 0.00 | -0.22 | Co3f2-O1 | -5.9% |
| Co3f2 | -0.17 | 0.00 | -0.22 | Co4f1-O1 | -0.2% |
| Co4f1 | 0.00 | 0.00 | -0.19 | Co4f1-O3 | -0.2% |
| Co4f2 | 0.00 | 0.00 | -0.19 | Co3f1-O3 | -5.9% |
| O1 | 0.00 | -0.06 | -0.05 |  |  |
| O2 | 0.00 | 0.06 | -0.05 |  |  |
| O3 | 0.00 | 0.08 | -0.05 |  |  |
| O4 | 0.00 | -0.08 | -0.05 |  |  |

The less dense (110)-B surface exposes only $Co^{3+}$ and $O^{2-}$ ions. All Co ions are equivalent and 4-fold coordinated to two surface and two second layer oxygens. There are two different types of surface oxygen ions: one is 2-fold (O-2f) coordinated to one surface Co ion and one 4-fold $Co^{2+}$ ion in the second layer; the other is 3-fold (O-3f) coordinated to one surface Co and two $Co^{3+}$ ions in the second layer (see Fig. 1). Table 3 shows the computed atomic relaxations for the (110)-B termination. The surface 2-fold and 3-fold oxygen ions behave differently upon relaxation: O-2f ions relax outwards and the bond with Co ions weakens, whereas O-3f ions relax inwards and their bonds to Co ion become stronger upon relaxation.

**Table 3.** Atomic displacements from bulk-like positions on the relaxed (110)-B surface. Displacements along the [001], [1$\bar{1}$0] and [110] directions are denoted as (Δx, Δy, Δz). Atoms are labeled as in Fig.1.

|  | Atomic displacement (Å) | | | Bond expansion | |
| --- | --- | --- | --- | --- | --- |
|  | Δx | Δy | Δz | Label | Δ |
| Co4f | -0.05 | 0.08 | -0.08 | Co1-O2f | 2% |
| O2f | -0.05 | -0.04 | 0.08 | Co1-O3f | -3% |
| O3f | 0.00 | -0.02 | -0.14 |  |  |

### B. Surface magnetization

In bulk $Co_3O_4$, only the $Co^{2+}$ ions at tetrahedral sites have a magnetic moment, whereas the $Co^{3+}$ ions at octahedral sites are non-magnetic. At the surface, the bulk symmetry is broken and the ionic coordinations are reduced, and therefore the magnetic properties of the surface cobalt ions can differ from those in the bulk. We computed the magnetic moments of the different surface ions on the (110)-A and (110)-B surfaces using a Löwdin charge analysis. The results, reported in Table 4, show that the surface $Co^{3+}$ ions are indeed magnetic on both terminations. Moreover, all surface Co ions have similar magnetic moments, which are also similar to the computed magnetic moment, $2.59\mu_B$, of the $Co^{2+}$ ions in bulk $Co_3O_4$.[8] Contour plots of the surface spin density for both terminations are shown in Fig. 4. We can see that on the (110)-A surface the oxygen ions are essentially non-magnetic, whereas on the (110)-B termination a slight spin polarization is present on the O-2f ions. The ionic magnetic moments in the second layer are already the same as in the bulk.

**Table 4.** Magnetic moments ($\mu_B$) of surface ions determined through Lowdin charge analysis

| A termination | | B Termination | |
|---|---|---|---|
| Ion type | Magnetic moment | Ion type | Magnetic moment |
| Co3f ($Co^{2+}$ in bulk) | 2.64 | Co4f ($Co^{3+}$ in bulk) | 2.56 |
| Co4f ($Co^{3+}$ in bulk) | 2.52 | O2f | 0.08 |
| O | 0.02 | O3f | 0.02 |

To determine the ground state surface magnetic configuration, we need to analyze the couplings between the different magnetic moments. In contrast to the bulk, where magnetic couplings are due to weak superexchange interactions (two metal ions separated by two oxygen ions), on the surface the presence of magnetic $Co^{3+}$ ions gives rise to normal superexchange interactions (two metal ions separated by one oxygen ion). There are normal superexchange interactions between surface Co ions, as well as between surface ions and the magnetic $Co^{2+}$ ions in the next layer (Fig.1). For the A termination, there are three different superexchange interactions. The coupling between surface neighboring Co-4f ions (J1 in Fig. 1) is via an intermediary oxygen ion in the second layer, with a Co-O-Co angle of 90°. According to the Goodenough-Kanamori-Anderson (GKA) rules,[32] the exchange interaction between them is ferromagnetic. The other two superexchange interactions are associated with angles of about 120°, for which the GKA rules do not make well defined predictions. The ground state ordering obtained by calculating the surface energies of different magnetic configurations is given in Table 5.

**Table 5**. Surface energies (meV/Å$^2$) of various magnetic configurations relative to the lowest energy state, taken as zero. Co4f ions are schematically indicated by underlined arrows, Co3f ions and Co$^{2+}$ ions in the second layer are indicated by arrows without underlines.

| A Termination | | B Termination | |
|---|---|---|---|
| ⇑↑⇑↑ <br> ↓ ↓ | 0.0 | ⇑ ⇑ <br> ↓ ↓ | 0.0 |
| ⇑↑⇑↑ <br> ↑ ↑ | 2.8 | ⇑ ⇑ <br> ↑ ↑ | 6.5 |
| ⇓↑⇓↑ <br> ↑ ↑ | 3.3 | | |

On the (110)-B termination, the distance between the surface magnetic Co-4f ions is quite large, and therefore the coupling between them can be considered weak. The only normal superexchange interaction is the one between surface Co-4f and Co$^{2+}$ ions in the second layer, which is also associated with a Co-O-Co angle of about 120°. From total energy differences between different magnetic configurations, it appears that this coupling is antiferromagnetic (see Table 5).

Based on the results in Table 5, the expected surface ground state magnetic configurations for the A and B terminations are schematically illustrated in Fig. 5. The surface region comprises the first and second layers, and is characterized by normal superexchange couplings, whereas below the second layer only weak antiferromagnetic superexchange interactions are present, as in bulk Co$_3$O$_4$. The presence of a ferrimagnetic surface region on the A termination is interesting. It can provide the mechanism to understand a number of experimental observations on Co$_3$O$_4$ nanostructures, notably: (i) the decoupling of magnetic core and shell contributions[16]; (ii) the ferrimagnetic behavior of porous nanostructures [17]; (iii) the exchange anisotropy phenomena observed in Co$_3$O$_4$ nanowires.[20]

## C. Surface electronic structure

Surface electronic states in the bulk band gap are of great interest because they can strongly influence the physical and chemical properties of semiconductor materials. For $Co_3O_4$, evidence of surface states in the band gap has been recently found in STM and STS studies on nanowires.[33] In this section we characterize the surface states on both $Co_3O_4$(110) terminations, by studying their energies and spatial distributions, i.e. on what ions these states are primarily localized, and how fast they decay when moving from the surface toward the bulk. The calculations were performed on symmetric slab models of 9 layers, for which spin densities are also symmetric, and spin up and spin down states are degenerate in energy. For this reason, we do not distinguish between spin up and spin down in the following; instead, all results include the sum over the two spin directions.

Figure 6 shows the computed band structures along various directions of the surface Brillouin zone. By comparison with the projected bulk structure (shaded area in Fig. 6), it is evident that on both surface terminations several surface state bands are present in the lower half of the bulk band gap. Partially occupied bands are present, indicating a metallic state. In Figure 7, we plot the Layer-Resolved Density of States (LRDOS) for surface models of A, B termination and a 4-layer bulk model. The DOS curves for the inner layers have a clear bulk-like character, as shown by the similarity between the bulk DOS and the DOS for the 4$^{th}$ and 5$^{th}$ layers of both surface models. At the surface new states appear close to the top of the valence band, while in the second layer, just below the surface, the tail of these states is still present, more prominent for the B termination, but starting from the third layer the DOS is already bulk-like.

To clarify the character of the surface states, in Figure 8 we show the partial densities of states, obtained by projecting the surface LRDOS onto the different surface oxygen and cobalt ions separately. On the (110)-A termination, surface states originate predominantly from surface O 2p states, and may be described as oxygen dangling bond-like states. On the (110)-B termination, both Co and oxygen contribute to the surface states which look more delocalized and metallic-like in character in comparison to those on the A termination. Partially metallic surface states are known to occur on other transition-metal oxide polar surfaces as well, notably on the Zn-terminated ZnO (000$\bar{1}$) surface,[34] suggesting that partial metallization may be a quite common phenomenon on surfaces of transition metal oxides.

Work functions for the two surface terminations were computed at both PBE and PBE+U levels. The results, reported in Table 6, clearly show a larger work function for the B termination relative to the A case, which can be attributed to the different surface dipoles on the two surfaces. We can also notice that PBE+U predicts a larger value of the work function in comparison to PBE, which may be attributed to the stabilization of the Co d states at the Fermi energy caused by the U term.

**Table 6**. Computed work functions (eV) from PBE and PBE+U calculations

|       | A Termination | B Termination |
| --- | --- | --- |
| PBE   | 3.96 | 4.59 |
| PBE+U | 5.28 | 5.97 |

**D. Compensating charges and bonding properties from the analysis of Wannier functions**

*D.1. Compensating charges*

A simple way to determine the value of the compensating charge for each termination is by calculating the total charge $Q_l$ in each layer of the slab. This can done very effectively and precisely by counting the number of Wannier centers (WCs) associated with each ion in that layer.[8] For the (110)-A termination, we find that the surface unit cell of the outermost layer has a total charge $Q_1 = +1$, instead of the value +2 found for the same layer in the bulk (see Figure 2). Similarly, for the (110)-B termination, the total charge of the top layer is $Q_1 = -1$, instead of the value -2 for the same layer in the bulk. Below the second layer, the charge of each layer is the same, +2 or -2, as in the bulk (Figure 2). As expected,[9] the compensating charges are $\Delta Q = -1$ and +1/cell for the A and B termination, respectively.

The same result can be also obtained by using a result of the modern theory of polarization[35] which shows that the compensating (or external) surface charge density $\sigma_{ext}$ is equal to the component of the bulk polarization, $P_{bulk}$, normal to the surface [35][36]

$$\sigma_{ext} = P_{bulk} \cdot \hat{n} \ . \qquad (1)$$

We determine $P_{bulk}$ from our previously calculated MLWFs and WCs for bulk $Co_3O_4$.[8] Eq. (1) then gives the surface charges on the A and B terminations simply using the frozen bulk ionic positions and ionic charges, without the necessity of slab calculations.

*D.2. Bonding properties*

For bulk $Co_3O_4$ different types of Wannier functions are present, namely $d$ states of $t_{2g}$ and $e_g$ symmetries localized on the cobalt ions, and Wannier functions with the character of $sp^3d$ bonds both between the cobalt and $O^{2-}$ ions.[8] These MLWFs show that the bonding character of $Co_3O_4$, although mainly ionic, has also a small covalent component.

As for the (110) surface, the MLWFs show that the surface is more covalent than the bulk, a result valid for both the A and B terminations. For instance, on the outermost surface layer there are several Wannier centers in mid position between different ions, see Fig. 9. The MLWF analysis also indicates that on the A termination the compensating excess electron is shared among two different $Co^{3+}$ ions, which are thus partially reduced. This compensating charge cannot be described by a single Wannier function or Kohn-Sham state. Similarly, on the B termination the compensating hole is shared between two $Co^{3+}$ ions which are thus are partially oxidized. On the B termination, one MLWF has relatively large spread, indicating that this termination has a metallic character.

*D.3. Non-symmetric stoichiometric slab models*

So far, our results were obtained from calculations on symmetric, non-stoichiometric slab models appropriate for the study of the surface properties of thick samples, on which charge compensation occurs naturally.[28] In the case of thin films and nanostructures, however, the polarity may remain uncompensated below a critical thickness[9] and possibly affect the properties and reactivity of these systems. It is therefore interesting to determine what is this critical thickness for $Co_3O_4$ (110). To this end we considered non-symmetric, stoichiometric slab models with different (even) number of layers and calculated the formation energy $E_{form}$ (total energy difference between the slab and an equal number of bulk $Co_3O_4$ units) and the electrostatic potential energy drop along the slab $\Delta V$ as a function of the number of layers. The results (Fig. 10) show that both $E_{form}$ and $\Delta V$ become approximately constant when the number of layers is larger than 4, implying that the critical thickness is 4 layers.

## IV SUMMARY AND CONCLUSION

We have presented an accurate and comprehensive computational study of the structural, electronic and magnetic properties of the polar $Co_3O_4$ (110) surface by the GGA+U method. We found the atomic relaxations give rise to a surface buckling of ~ 0.2 Å on both surface terminations. Surface energy calculations indicate that the (110)-A termination is more stable in a wide range of the oxygen chemical potential, in agreement with surface science experiments.[14]

The $Co^{3+}$ ions do not have a magnetic moment in the bulk but become magnetic at the surface which leads to interesting surface magnetic properties, as found also in recent experiments on $Co_3O_4$ nanostructures.[16,17,19,20] From band structure and density of states calculations, we found that surface electronic states are present in the bulk band gap for both terminations, consistent with STM experiments on $Co_3O_4$ nanowires.[33] The B termination is found to have a more pronounced metallic character compared to the (110)-A surface. It has also a larger work function, which could play an important role in the study of surface redox reactions. Maximally localized Wannier functions clearly show that charge compensation takes place on the top layer of both terminations. They also reveal that the surface is more covalent with respect to the bulk. Calculations on asymmetric models predict a critical thickness for polarity compensation of 4 layers. We hope that these predictions can be tested experimentally in the near future.

**Acknowledgement**

This work was supported by DoE-BES, Division of Materials Sciences and Engineering under Award DE-FG02-06ER-46344, and Division of Chemical Sciences, Geosciences and Biosciences under Award DE-FG02-05ER15702. We used resources of the National Energy Research Scientific Computing Center (DoE Contract No. DE-AC02-05CH11231) and Center for Nanoscale Materials, supported by the U. S. Department of Energy, Office of Science, Office of Basic Energy Sciences (DoE Contract No. DE-AC02-06CH11357). We also acknowledge use of the TIGRESS high performance computer center at Princeton University which is jointly supported by the Princeton Institute for Computational Science and Engineering and the Princeton University Office of Information Technology.

**Figures and Captions**

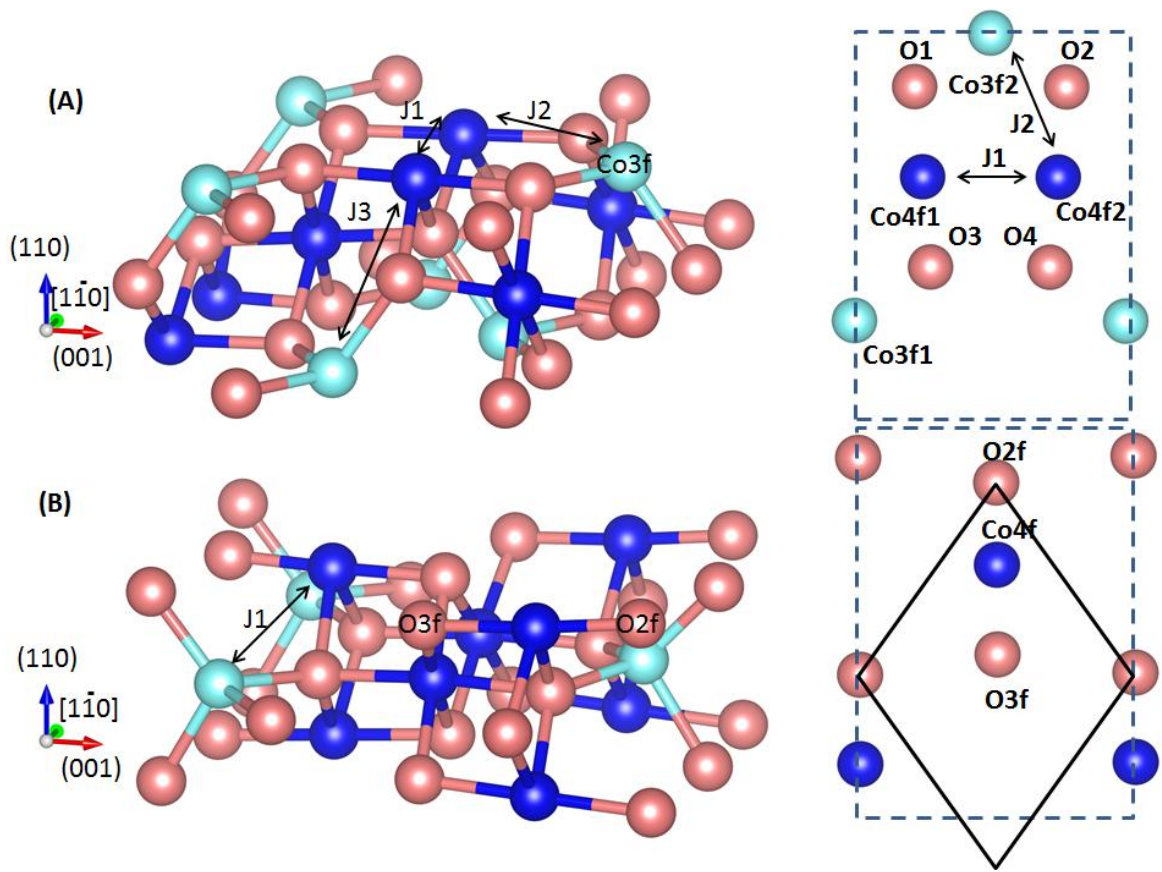

**Figure 1.** Ball and stick models of the A(top) and B (bottom) terminations of $Co_3O_4(110)$. Left: side views. Right: top views (surface layer only). Superexchange interactions between surface Co ions are indicated. Dashed lines denote a rectangular cell which is the primitive surface cell for the (110)-A terminations and a surface cell twice the primitive cell for the B-(110) termination; the solid line indicates the primitive cell of the B termination. Light cyan and navy blue balls indicate $Co^{2+}$ and $Co^{3+}$ ions, red ones indicate $O^{2-}$ ions.

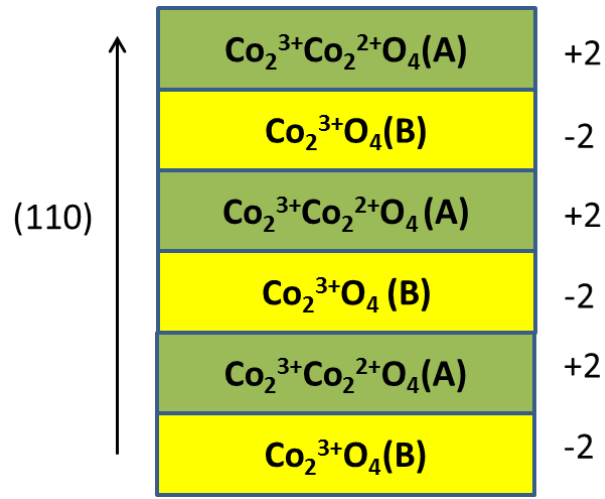

**Figure 2.** Sketch of a $Co_3O_4$(110) slab model as a stack of charged layers.

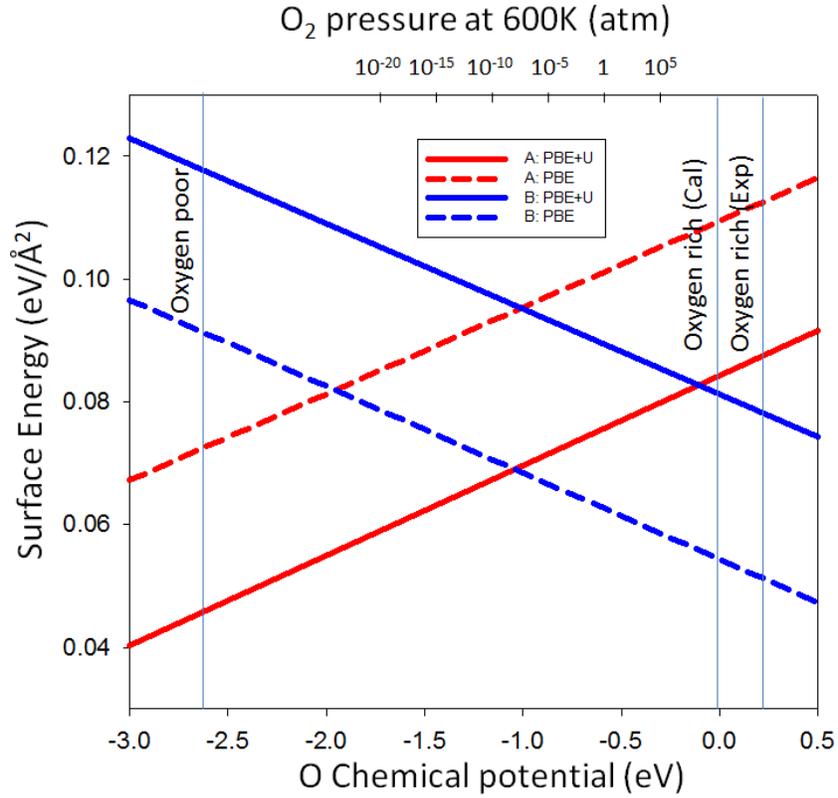

**Figure 3:** Surface energies of the (110)-A and (110)-B surfaces from PBE and PBE+U calculations. Vertical lines define the allowed range of the oxygen chemical potential $\mu_O' \equiv \mu_O - \frac{1}{2} E_{tot}(O_2)$: the leftmost line indicates the oxygen-poor limit, while the lines on the right indicate the oxygen rich limit determined using the computed ($\mu_O' = 0$ line) and experimental (rightmost line) $O_2$ binding energy, respectively.

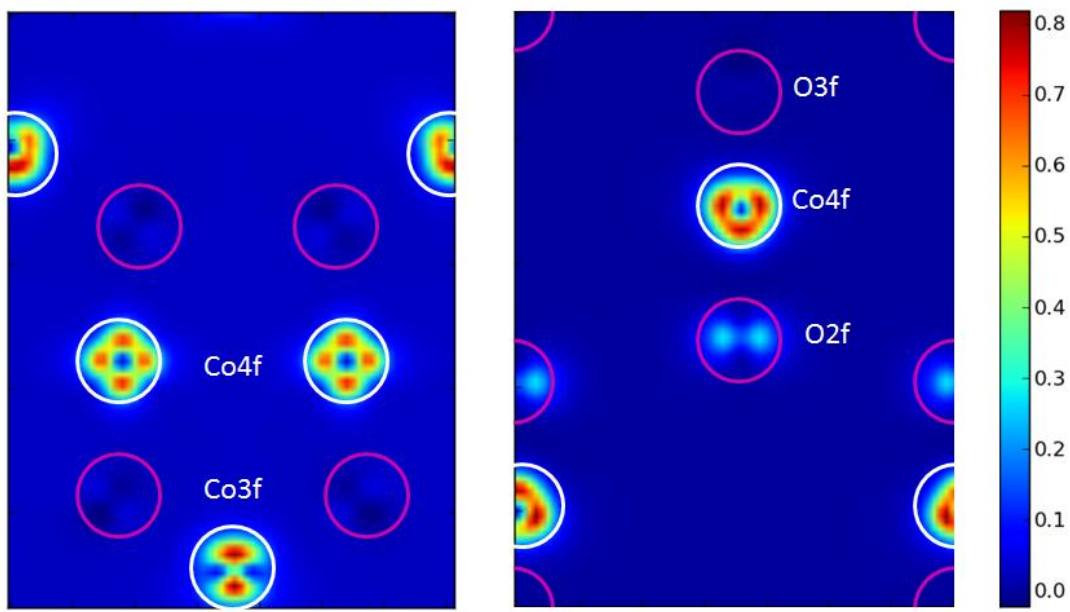

**Figure 4**. Contour plots of the surface spin density on the (110)-A (left) and (110)-B (right) surfaces. The scale in the bottom has units of $\mu_B$. The positions of the Co ions are indicated by white circles and those of the oxygen ions by red circles.

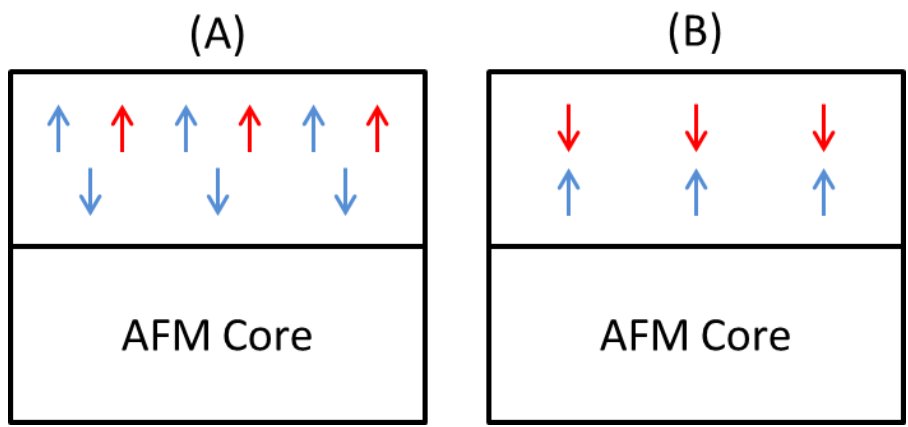

**Figure 5.** Magnetic ground state configurations of the (110)-A (left) and (110)-B (right) surfaces, as inferred from the surface energies in Table 5. Red (blue) arrows refer to Co4f (Co3f and $Co^{2+}$ in second layer) ions.

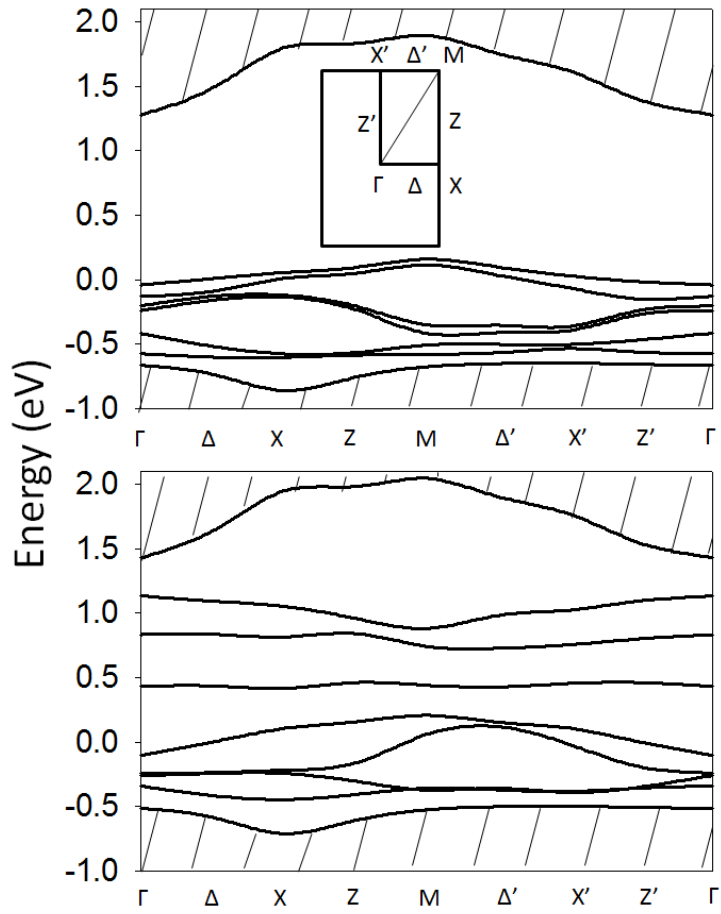

**Figure 6**. Band structures for symmetric slabs of 9 layers terminated by (110)-A (top) and (110)-B (bottom) surfaces. Spin up and spin down states are degenerate in energy (see text). The shaded area represents the projected bulk bands. The energy zero corresponds to the Fermi energy. For both terminations partially occupied bands are present, indicating that the surfaces are metallic.

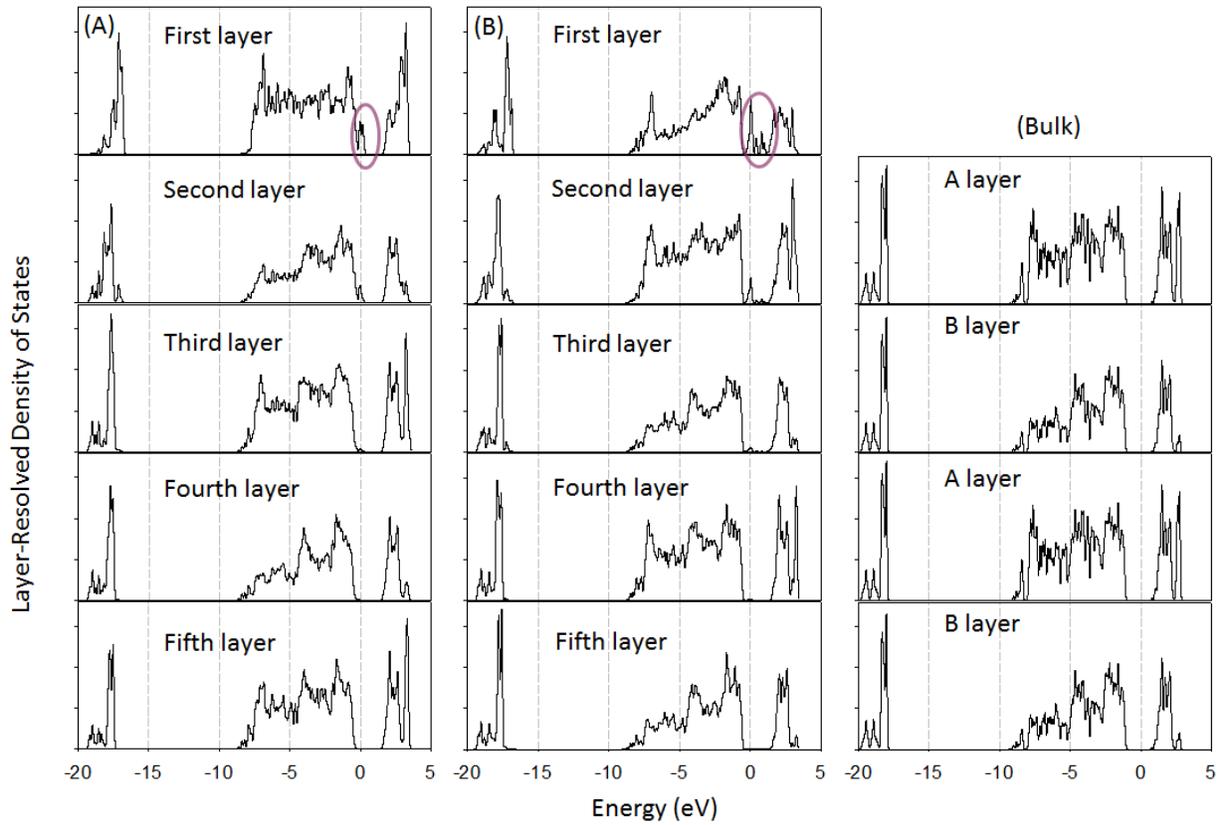

**Figure 7.** Spin-averaged layer-resolved Density of States for the (110)-A (left), (110)-B (middle) surfaces and bulk (right) of $Co_3O_4$. Surface states in the surface layer are highlighted. The energy zero corresponds to the Fermi energy. The DOS curves for the inner layers in the slab calculations have a clear bulk-like character, as shown by the similarity between the bulk DOS and the DOS for the 4$^{th}$ and 5$^{th}$ layers of both surface models.

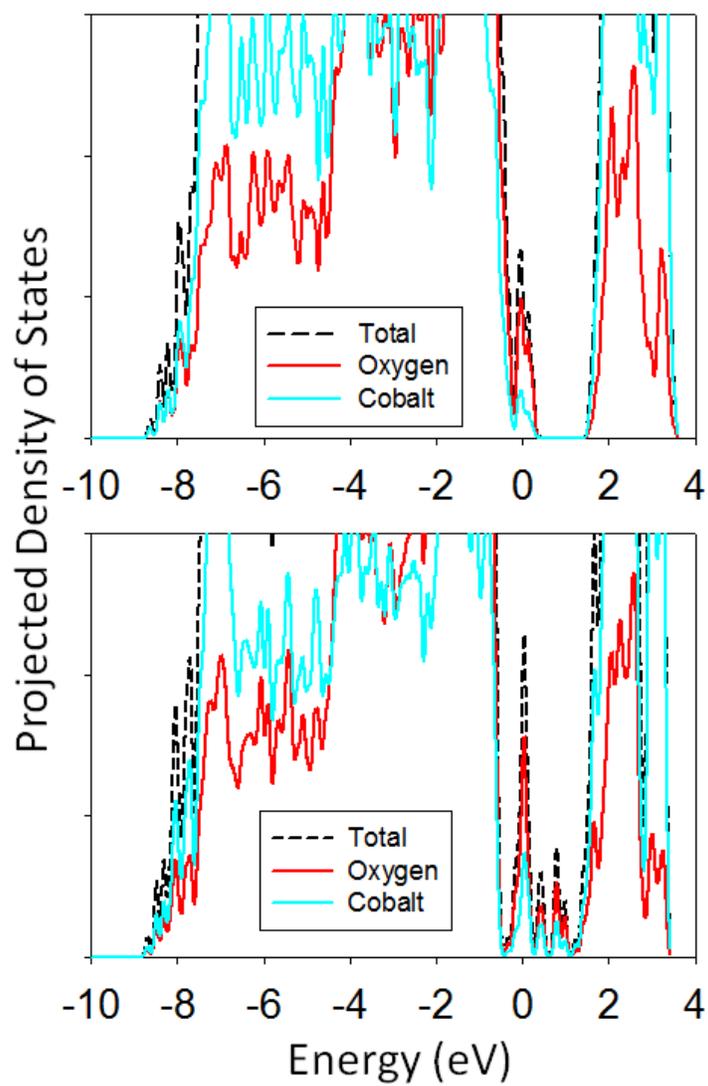

**Figure 8**. Spin-averaged projected density of states on the (110)-A (top) and (110)-B (bottom) surfaces. The energy zero corresponds to the Fermi energy.

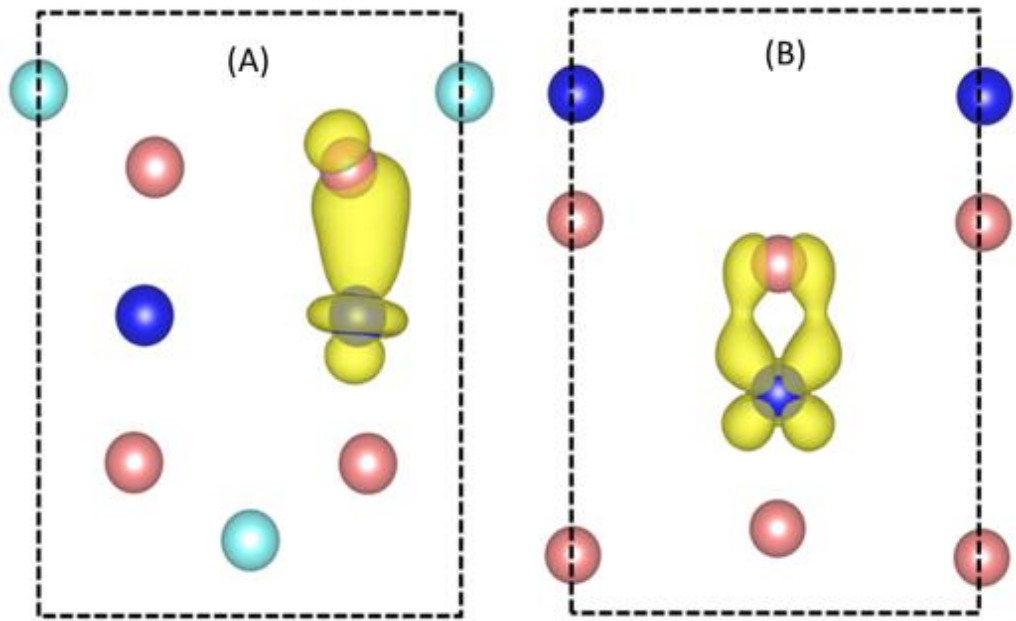

**Figure 9**. Charge densities of typical covalent MLWFs on the (110)-A (left) and (110)-B (right) termination.

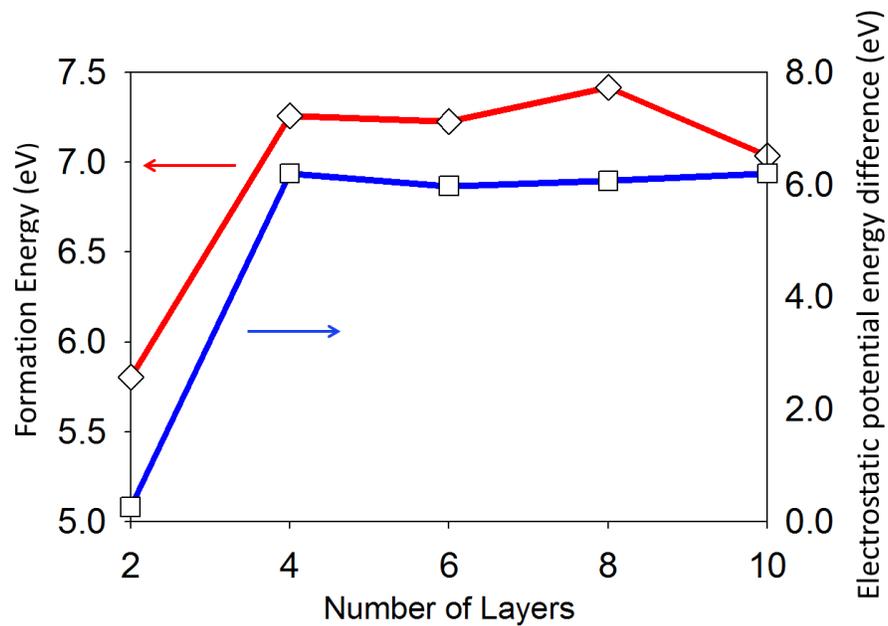

**Figure 10.** Formation energy and electrostatic potential energy drop (eV) for stoichiometric slab models as a function of the number of layers in the slab.